\documentclass[prx,aps,twocolumn,showpacs,longbibliography]{revtex4-1}
\usepackage{graphicx}
\usepackage{amsmath}
\usepackage{bm}
\usepackage{amstext}
\usepackage{amsxtra}
\usepackage{float}
\usepackage{upgreek}

\usepackage{times}

\usepackage[usenames, dvipsnames]{color}

\begin{document}

\title{Observation of Weak Collapse in a Bose-Einstein Condensate}

\author{Christoph Eigen, Alexander L. Gaunt, Aziza Suleymanzade, Nir Navon, Zoran Hadzibabic, and Robert P. Smith}
\affiliation{Cavendish Laboratory, University of Cambridge, J. J. Thomson Avenue, Cambridge CB3 0HE, United Kingdom }

\begin{abstract}

We study the collapse of an attractive atomic Bose-Einstein condensate prepared in the uniform potential of an optical-box trap.
We characterise the critical point for collapse
and the collapse dynamics, observing universal behaviour in agreement with theoretical expectations.
Most importantly, we observe a clear experimental signature of the counterintuitive weak collapse, namely that making the system more unstable can result in a smaller particle loss.
We experimentally determine the scaling laws that govern the weak-collapse atom loss, providing a benchmark for the general theories of nonlinear wave phenomena.

\end{abstract}

\date{\today}

\pacs{ 03.75.Nt, 03.75.Kk, 67.85.-d, 67.85.Hj}


\maketitle

\section{Introduction}

Wave collapse occurs in a wide range of physical contexts, including optics, atomic and condensed-matter physics.
Generally, collapse occurs if an attractive nonlinearity exceeds a critical value.
If the collapse is triggered at time $t=0$,  the wave amplitude asymptotically diverges at some point in space as the collapse time $t_{\rm c}$ is approached.
In practice, the amplitude divergence results in dissipation of wave energy (or particle loss).

The unifying theoretical framework for understanding different collapse phenomena is provided by the nonlinear Schr{\"o}dinger equation, which has been extensively studied for various forms of nonlinearity~\cite{Sulem:1999, Fibich:2015}.
This general formalism is applied to self-focusing of light~\cite{Askaryan:1962, Chiao:1964, Kelly:1965, Pilipetskii:1965, Hercher:1964}, collapse of Langmuir waves~\cite{Zakharov:1972,Wong:1984} and Bose-Einstein condensates (BECs)~\cite{Ruprecht:1995, Kagan:1996a, Kagan:1997, Kagan:1998, Eleftheriou:2000}, and even surface water waves \cite{Davey:1974,Papanicolaou:1994}.

\begin{figure} [btp]
\centering
\includegraphics[width=1\columnwidth]{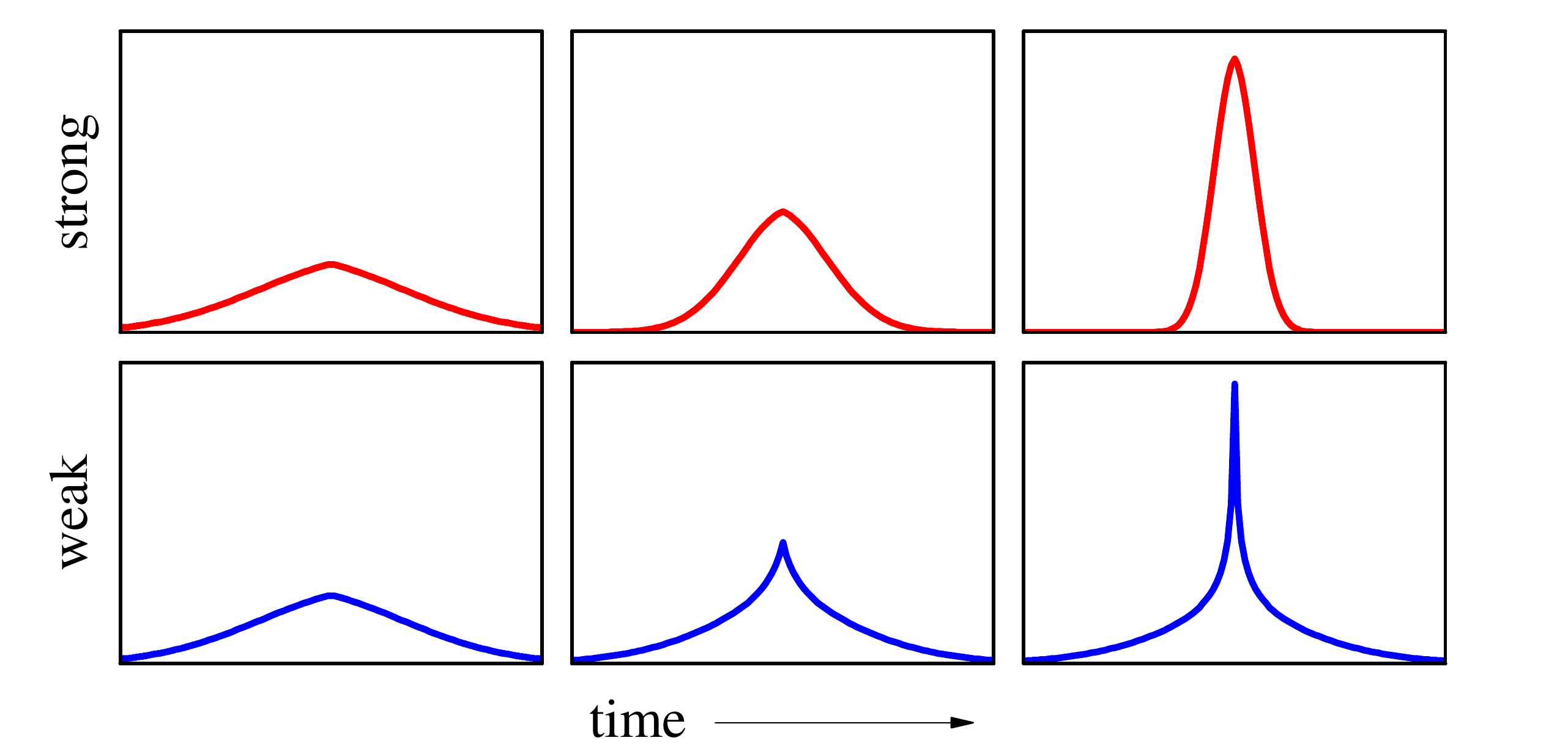}
\caption{Strong versus weak collapse (cartoon). In strong collapse, a finite portion of the wave (here 100\%, for simplicity) collapses towards the singularity.
In weak collapse, as time progresses, a diminishing fraction of the wave approaches the singularity, with long tails left behind.
}
\label{fig:Cartoon}
\end{figure}

In this framework, wave collapse is classified as either {\it strong} or {\it weak} (see Fig.~\ref{fig:Cartoon}). In a strong collapse, a finite fraction of the wave collapses towards the singularity.
On the other hand, in a weak collapse~\cite{Zakharov:1975,Zakharov:1985,Zakharov:1986} the fraction of the wave that (in absence of dissipation) ultimately reaches the singularity vanishes.
This has the counterintuitive practical implication that making the system {\it more unstable}, by quenching the nonlinearity further beyond the critical point, can result in {\it less} dissipation~\cite{Zakharov:1986, Berge:2002}.
Qualitatively, once the collapse is triggered, for stronger attractive interactions it happens faster and progresses further before dissipative processes halt it; consequently the wave amplitude is larger at the point in time when dissipation occurs, and for weak collapse this means that a smaller fraction of the wave is actually dissipated. 
To our knowledge, weak collapse has not been experimentally observed in any physical system.

An atomic BEC with $s$-wave two-body
interactions is modelled by the Gross-Pitaevskii (GP) equation, with a cubic nonlinearity proportional to the scattering length $a$, which can be dynamically tuned via a Feshbach resonance~\cite{Chin:2008}.
The BEC is prone to collapse for any $a<0$,
but a kinetic-energy barrier makes it metastable up to a critical interaction strength~\cite{Ruprecht:1995, Kagan:1996a, Kagan:1997, Kagan:1998, Eleftheriou:2000}. If the BEC becomes unstable and collapses, dissipation occurs through three-body recombination that results in particle loss.
Importantly, the three-dimensional GP equation is expected to provide an example of weak collapse.

Previous collapse experiments with atomic BECs~\cite{Gerton:2000, Roberts:2001, Donley:2001, Cornish:2006, Altin:2011, Compton:2012} (see also~\cite{CollapseFootnote1})
were performed in the traditional setting of a harmonic trap.
The critical point~\cite{Roberts:2001} and collapse times~\cite{Donley:2001, Altin:2011} were in general agreement with theoretical expectations~\cite{Ruprecht:1995, Santos:2002, Saito:2002, Adhikari:2002, Savage:2003, Metens:2003,Ueda:2003, Carr:2004,Wuster:2005}, but no evidence of weak collapse was observed; the atom loss was only seen to grow with $|a|$~\cite{Cornish:2006}.

In this article we study BEC collapse in a new experimental setting, using a $^{39}$K condensate~\cite{Roati:2007,Campbell:2010} prepared in the uniform potential of an optical-box trap~\cite{Gaunt:2013}; for details of our setup see Appendix A.
The combination of large system size (up to $41\,\mu$m) and fine tuning of the scattering length (with a resolution of $0.03\,a_0$, where $a_0$ is the Bohr radius)
gives us a very large dynamic range: we observe metastable attractive BECs with up to $2\times 10^5$ atoms, and collapse times that vary between $3$ and $400$~ms. We demonstrate the expected scaling of the critical scattering length $a_{\rm c}$ with the BEC atom number $N$ and the system size $L$, and show that the collapse time can be expressed as a universal function of the dimensionless interaction strength $aN/L$.
Most importantly, we observe conclusive evidence for weak collapse, namely the counterintuitive decrease of the atom loss with increasing $|a|$, and experimentally determine the scaling laws that govern the weak-collapse atom loss.
The weak nature of the collapse is directly revealed only by resolving single collapse events, and is obscured in the multiple collapse regime, which has been seen in previous cold atom experiments.

In Sections~\ref{sec:CriticalPoint}-\ref{sec:WeakCollapse} we address in turn: (i) the {\it critical point} for the collapse, (ii) the {\it collapse dynamics} in a system that is suddenly made unstable by an interaction quench, and (iii) the {\it  aftermath}
of the collapse, which reveals its weak nature.

\section{Critical Point}
\label{sec:CriticalPoint}

The starting point for our discussion is the GP equation for a homogeneous box potential, with a heuristically added three-body loss term~\cite{Kagan:1998}:
\begin{equation}
i \hbar \frac{\partial \psi}{\partial t}=-\frac{\hbar^2}{2m}\nabla^2\psi+ \frac{4\pi \hbar^2 a}{m}  |\psi|^2\psi - i \frac{\hbar K_3 }{2}|\psi|^4\psi \, ,
\label{eq:gp}
\end{equation}
where $m$ is the atom mass, $K_3$ is the three-body loss coefficient,
$\psi$ is normalised to the atom number $N$,
and the boundary condition is  $\psi = 0$ at the trap walls.

We use a cylindrical box trap~\cite{Gaunt:2013} of variable length $L$ and radius $R$, and always set $R=L/2$, so $L$ is the only lengthscale characterising the system size.
We may thus rewrite Eq.~(\ref{eq:gp}) in a dimensionless form, defining $\mathbf{\tilde{r}} = \mathbf{r}/L$  and $\tilde{t} = t / \tau_0$,  with characteristic time $\tau_0=2m L^2/\hbar$:
\begin{equation}
i \frac{\partial \tilde{\psi}}{\partial \tilde{t}}=-\nabla^2\tilde{\psi}+\alpha|\tilde{\psi}|^2\tilde{\psi}-i\eta |\tilde{\psi}|^4\tilde{\psi} \, ,
\label{eq:rgp}
\end{equation}
where
\begin{equation}
\alpha= \frac{8\pi a N}{L}\quad \textrm{and} \quad \eta= \frac{N^2 mK_3}{\hbar L^4} \, ,
\label{eq:dimcoef}
\end{equation}
and $\tilde{\psi}$ is initially normalised to unity.
For the range of scattering lengths that we study, we assume that $K_3$ is constant \cite{Shotan:2014, Lepoutre:2016}, with the value $1.3 (5) \times 10^{-41}$\,m$^{6}$s$^{-1}$~\cite{Fattori:2008}.
The corresponding value of $\eta$, for all our $L$ and $N$ values, is very small ($<6 \times 10^{-4}$) and thus three-body loss is negligible in our (meta)stable condensates.
However, if the BEC collapses, significant loss occurs, providing the primary experimental signature of the collapse.

Neglecting the atom loss in a metastable BEC, based on Eq.~(\ref{eq:rgp}) the critical interaction strength, $\alpha_{\rm c}$, can depend only on the boundary conditions, {\it i.e.} the box shape. For a family of self-similar boxes ($R/L=$~const.) it should be a universal constant, so $a_{\rm c} \propto L/N$.

To experimentally  study the critical point for collapse, we prepare a stable BEC at $4\,a_0$, then over $1$~s ramp the scattering length to a variable $a<0$, and wait for $2$~s before turning off the trap and imaging the atoms  after $100$~ms of time-of-flight (ToF) expansion.  We image the cloud along the axial direction of our cylindrical trap, and for ToF we jump the scattering length to  $20\,a_0$.

\begin{figure} [tbp]
\centering
\includegraphics[width=1\columnwidth]{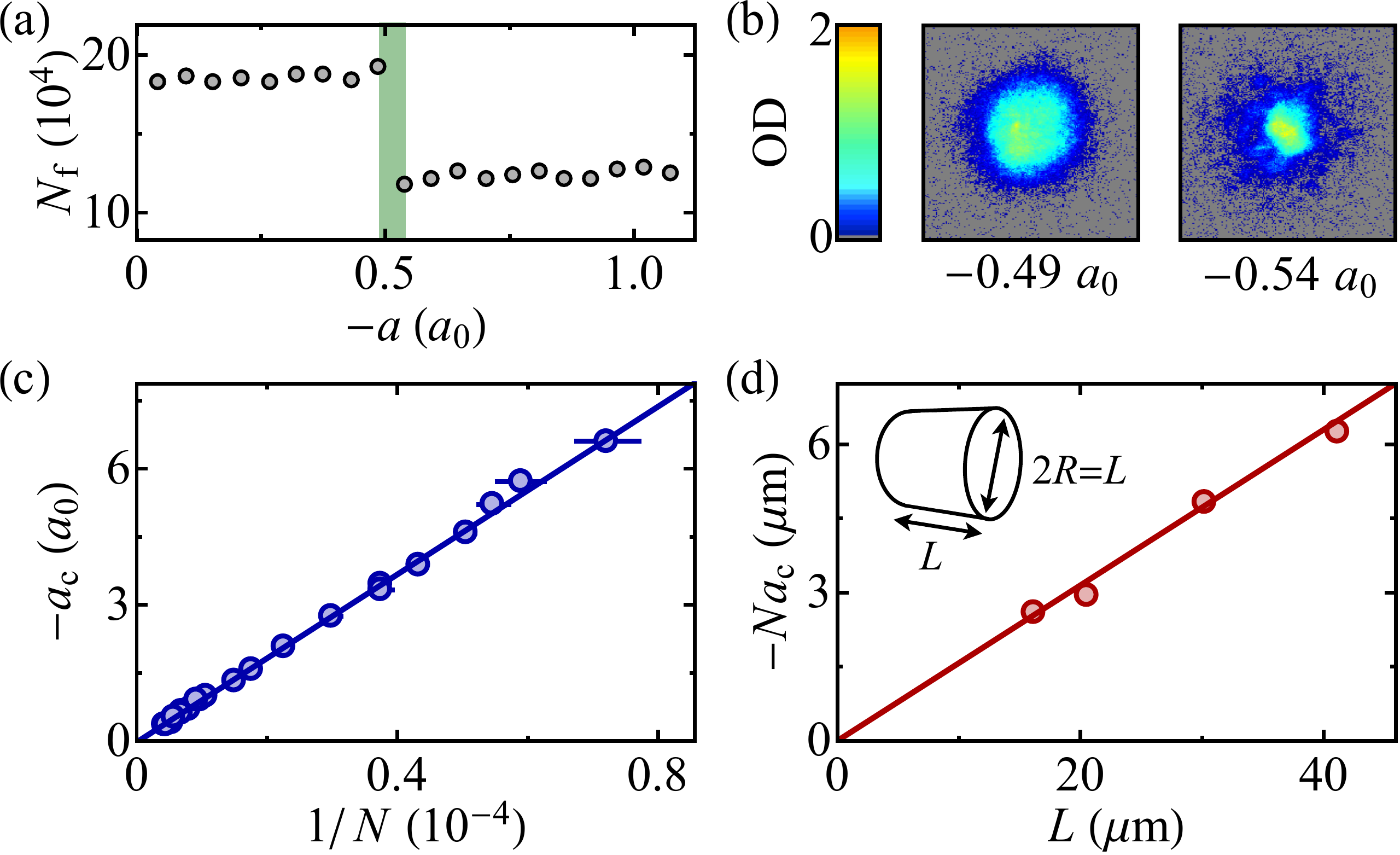}
\caption{Critical point for collapse.
(a) Final atom number, $N_{\rm f}$, after ramping to a variable negative $a$, for  $L=30(1)~\mu$m and initial $N = 18.7(5) \times 10^4$. The critical scattering length $a_{\rm c}$ is seen in the sharp drop in $N_{\rm f}$.
(b) ToF images taken on either side of the critical point.
(c) Variation of  $a_{\rm c}$ with $N$, for $L=30~\mu$m. The linear fit confirms the expected scaling $a_{\rm c} \propto 1/N$. (d) Variation of $Na_{\rm c}$ with $L$. The linear fit confirms the scaling $N a_{\rm c} \propto L$.
}
\label{fig:ac}
\end{figure}

In Fig.~\ref{fig:ac}(a) we show how, for a given initial $N$, the final atom number depends on the negative $a$. A well defined $a_{\rm c}$ is signaled by a sharp drop in the atom number. As shown in Fig.~\ref{fig:ac}(b), the atom loss is accompanied by a qualitative change in the appearance of the cloud in ToF.

In Fig.~\ref{fig:ac}(c) we plot $a_{\rm c}$ for $ L=30$\,$\mu$m and a wide range of $N$ values, from $10^4$ to $2\times 10^5$.
We clearly observe the expected scaling  $a_{\rm c} \propto 1/N$ (see also Appendix B).
In Fig.~\ref{fig:ac}(d) we plot the measured  $N a_{\rm c}$ versus box size and confirm the scaling $Na_{\rm c} \propto L$.
We find that the dimensionless critical interaction strength is $\alpha_{\rm c}= -  4(1)$,  where the error includes the systematic uncertainties in box size and absolute atom number calibration.  For comparison, numerical simulations of the GP equation for our box geometry give  $\alpha_{\rm c}= - 4.3$.

\section{Collapse Dynamics}
\label{sec:Dynamics}

To study the collapse dynamics, we perform interaction-quench experiments~\cite{Donley:2001}.  We prepare  a BEC just above $a_{\rm c}$ and then quench the scattering length to a variable $a < a_{\rm c}$ to initiate the collapse
(see Appendix B for more details).
After a variable hold time $t$ we jump the scattering length to $20\,a_0$, switch off the trap, and observe the cloud in ToF.

\begin{figure}[t!]
\centering
\includegraphics[width=1\columnwidth]{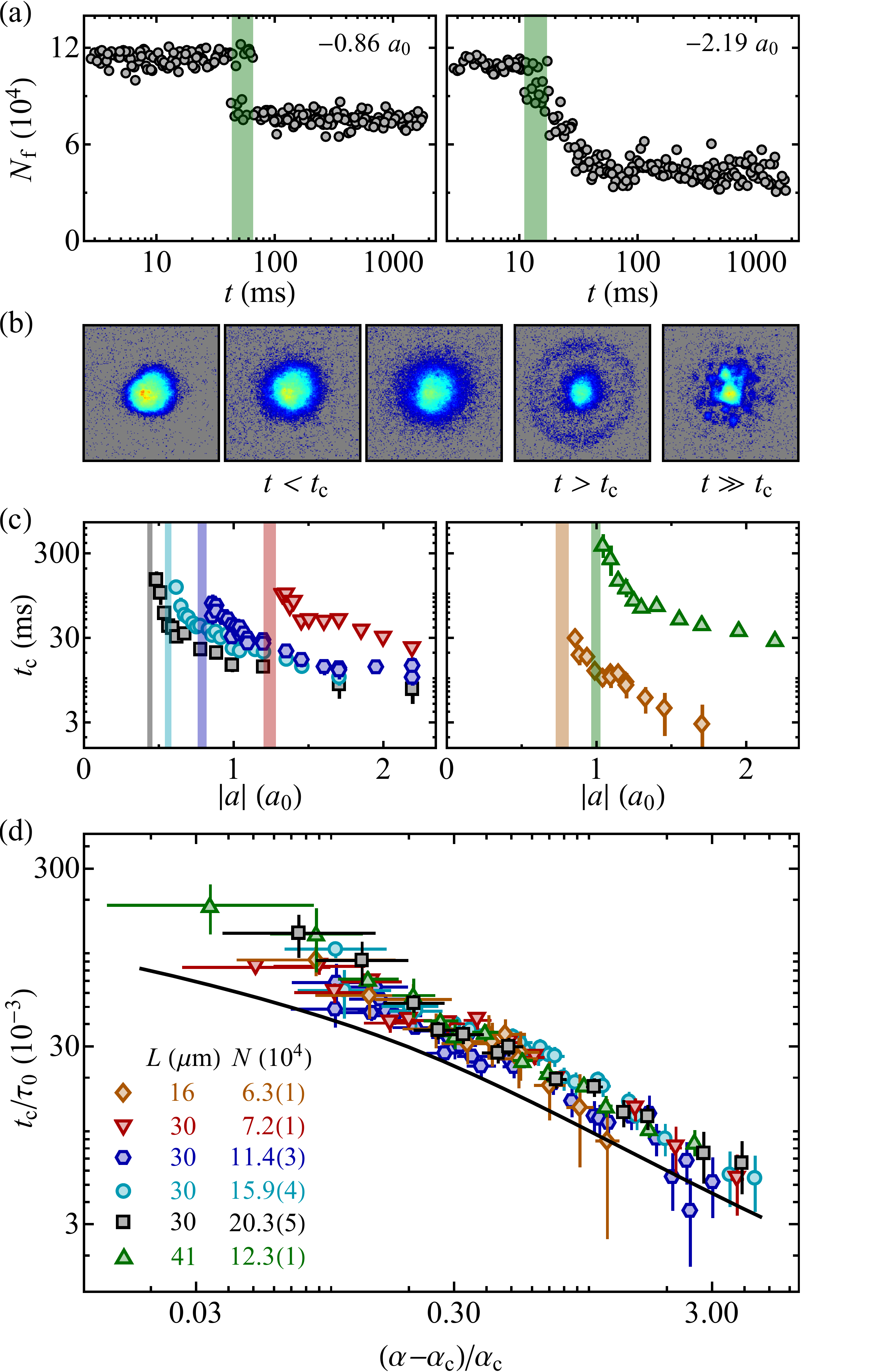}
\caption{Collapse dynamics.
(a) Atom number versus time after quenches to $a=-0.86\,a_0$ (left) and $-2.19\,a_0$ (right); here $L=30\,\mu$m and $N=11.4 \times10^4$, corresponding to $a_{\rm c} = -0.79\,a_0$. Green bands indicate $t_{\rm c}$ and its uncertainty.
(b) Typical ToF images at various stages after the quench (here for $a=-1.02\,a_0$).
(c) Collapse time versus $a$ for six data sets taken for various $L$ and $N$; see legend in (d). The shaded bands indicate $a_{\rm c}$  values. (d) Universal collapse dynamics. We plot the dimensionless collapse time versus the reduced distance from the critical point, for all six data sets. The solid line shows the results of lossless GP simulations without any free parameters.
}
\label{fig:jump}
\end{figure}

As shown in the left panel of Fig.~\ref{fig:jump}(a), for quenches close to the critical point (small $|a-a_{\rm c}|$),  at $t_{\rm c}$ the atom number suddenly drops to a stable lower value.  We understand this as a single collapse event.  On the other hand, for large quenches [right panel of Fig.~\ref{fig:jump}(a)], the atom number appears to gradually decay until it stabilises.
Such behaviour, also seen in~\cite{Donley:2001, Altin:2011}, is understood as arising from a series of multiple (experimentally unresolved) collapses~\cite{Malkin:1988, Vlasov:1989, Zakharov:1989, Kagan:1998, Saito:2001a, Saito:2001b, Saito:2002, Santos:2002, Berge:2002}, and we accordingly associate $t_{\rm c}$ with the onset of the atom loss \cite{Definetc}.
(For further evidence for the occurrence of single and multiple discrete collapse events see Appendix C.)

In Fig.~\ref{fig:jump}(b) we show typical ToF images for different times after the quench. At $t < t_{\rm c}$, before any change in the atom number occurs, the swelling of the cloud in ToF reveals the shrinking of the wavefunction in-trap.
Right after $t_{\rm c}$, within the first ${\approx}\,10$~ms, we observe that the remnant cloud consists of a lower-energy central part and a higher-energy shell, reminiscent of the atom bursts generated during collapse in \cite{Donley:2001}. At   longer times we observe more irregular patterns.
We see a similar shell structure in images taken along a perpendicular direction, which implies that the outgoing atom shell is spherical. Based on its size in ToF, the shell expands at a rate of ${\approx}\,2$~mm/s, which is consistent with it reflecting off the trap walls and interfering with the central part of the cloud after ${\approx}\,10$~ms.

In Fig.~\ref{fig:jump}(c) we plot $t_{\rm c}$ versus $a$ for six data sets taken with different $L$ and $N$ values. We observe $t_{\rm c}$ values
that vary between 3 and $400\,$ms.
In Fig.~\ref{fig:jump}(d) we show that all the data points fall onto a single universal curve if we plot the dimensionless collapse time, $t_{\rm c}/\tau_0$, versus the reduced distance from the critical point,  $(a-a_{\rm c})/a_{\rm c} \equiv (\alpha-\alpha_{\rm c})/\alpha_{\rm c}$.
In general, $t_{\rm c}$ could also depend on $\eta$, but the universal behaviour seen in Fig.~\ref{fig:jump}(d) shows that this effect is negligible for our range of $\eta$, between $4 \times 10^{-5}$ and $4 \times 10^{-4}$.
The solid line in Fig.~\ref{fig:jump}(d) shows results of lossless GP simulations, without any free parameters; we reproduce a very similar dependence of $t_{\rm c}$ on $\alpha$, although the numerical values are systematically slightly below the experimental ones.

\section{Weak Collapse}
\label{sec:WeakCollapse}

\begin{figure} [btp]
\centering
\includegraphics[width=1\columnwidth]{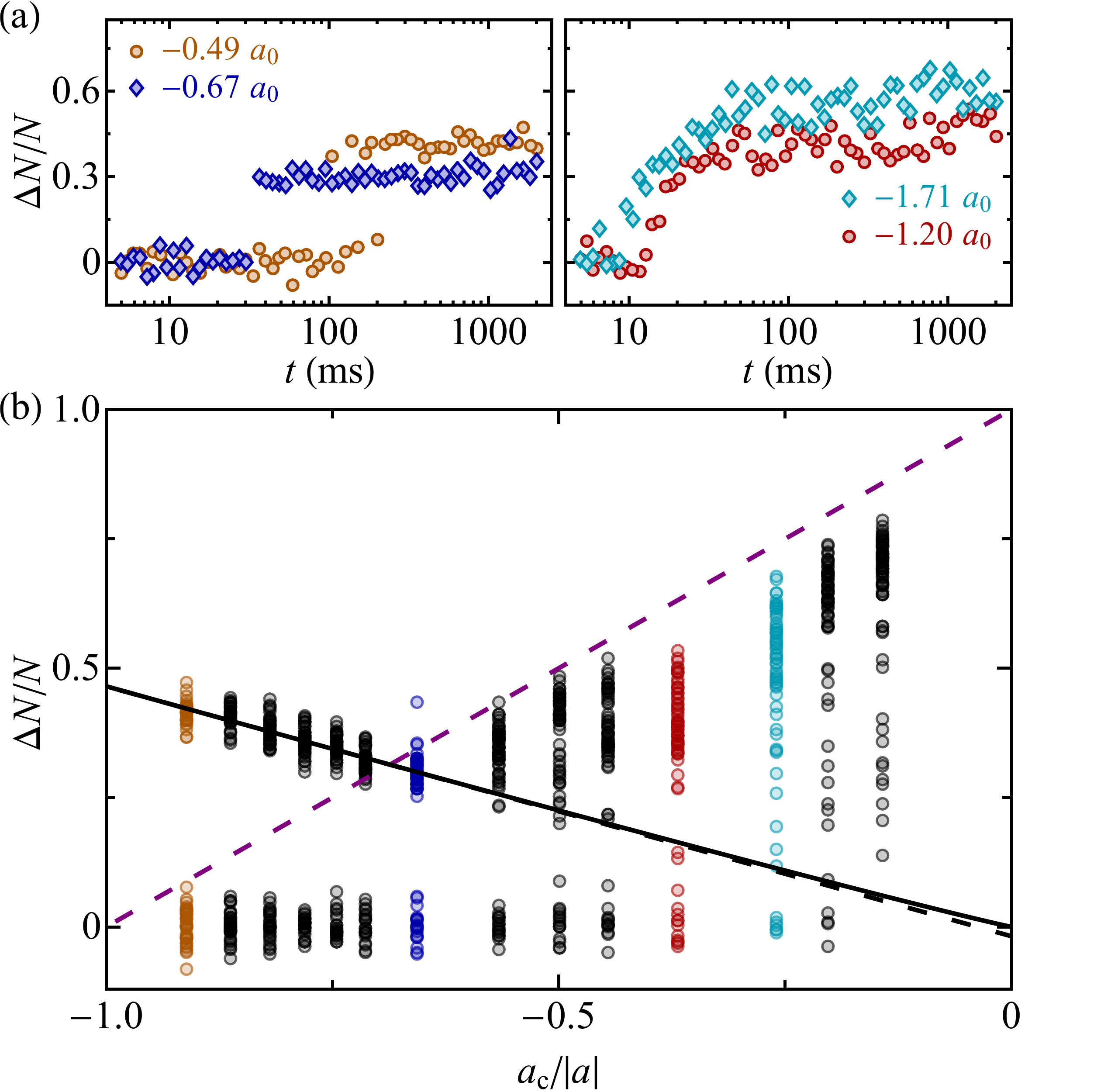}
\caption{Observation of weak collapse. Here $L=30\,\mu$m and $N=20.3 \times10^4$, corresponding to $a_{\rm c} = -0.44\,a_0$.
(a) Atom loss versus time after quenches to various $a$ values. For small $|a- a_{\rm c}|$ (left panel) $\Delta N/N$ decreases with increasing $|a|$,
as expected for a weak collapse, while
for large $| a- a_{\rm c}|$ (right panel) the opposite trend is seen.
(b) Summary of atom loss for all $t$ and $a<a_{\rm c}$. The coloured points are the data shown in (a), and the points clustered around $\Delta N =0$ correspond to $t< t_{\rm c}$.
Single-collapse loss monotonically decreases with increasing $|a|$ and extrapolates to zero for $a_{\rm c}/|a| \rightarrow 0$ (dot-dashed and solid black lines, see text),
 confirming the prediction of the weak-collapse theory.
For $a_{\rm c}/|a| > - 0.6$ single collapse does not re-stabilise the system and multiple collapse occurs. The dashed purple line shows the equilibrium stability criterion,
$\Delta N/N = 1 - a_{\rm c}/a$ (see text).
}
\label{fig:Nlost}
\end{figure}

We now turn to the aftermath of the collapse.
Since $\tilde{\psi}$ is initially normalised to unity, the {\it fractional} atom loss, $\Delta N/N$, should be some universal function of $\alpha$ and $\eta$; here $\Delta N = N-N_{\rm f}$ is the difference between the initial (pre-collapse) and the final (time-dependent) atom number.
The counterintuitive implication of the weak-collapse theory is that $\Delta N/N$ {\it decreases} if the BEC is made {\it more unstable}, by quenching $a$ to a more negative value.

In Fig.~\ref{fig:Nlost} we focus on one data set, for fixed $L=30\,\mu$m and $N=20.3 \times10^4$.
As we illustrate in the left panel of Fig.~\ref{fig:Nlost}(a), close to the critical point, where we observe only single-collapse events, the atom loss indeed decreases with increasing $|a|$, indicating weak collapse.
On the other hand, as shown in the right panel of Fig.~\ref{fig:Nlost}(a), in the regime of large quenches and multiple collapse, the atom loss in the long-time limit shows the opposite trend; only this type of behaviour was seen in harmonic-trap experiments~\cite{Donley:2001, Cornish:2006}.

In Fig.~\ref{fig:Nlost}(b) we present a consistent picture of the atom-loss trends for all $a< a_{\rm c}$, from $a/a_{\rm c} \approx 1$ to $a/a_{\rm c} \rightarrow \infty$.
Here we plot $\Delta N/N$ versus $a_{\rm c}/|a|$, and for each $a$ show $\Delta N/N$ values observed for all $t$;
the points clustered around $\Delta N=0$ correspond to $t<t_{\rm c}$.

The single-collapse regime, $a_{\rm c}/|a| < - 0.6$, is clearly identified by the small spread of the non-zero $\Delta N$ values.
The single-collapse atom loss clearly decreases with increasing $|a|$, and extrapolates to zero for $a_{\rm c}/|a| \rightarrow 0$.  This is the unambiguous signature of a weak collapse. The dot-dashed black line shows a linear extrapolation, which gives $\Delta N/N = -0.02(2)$ for $a_{\rm c}/|a| = 0$, while the (almost indistinguishable) solid black line shows a power-law fit,
$\Delta N/N \propto |a|^{-1.05(7)}$.

For $a_{\rm c}/|a| > - 0.6$, multiple collapse occurs, because the diminishing single-collapse atom loss does not re-stabilise the system.
However, we see that even in this regime the minimal loss we observe at each $a$ still follows the weak-collapse trend (solid black line).
It is also instructive to plot the function $\Delta N/N = 1 - a_{\rm c}/a$ (dashed purple line); this is atom loss such that,  after a quench to a given $a < a_{\rm c}$, the atom number drops to the new critical value $N_{\rm c} (a) = \alpha_{\rm c} L /(8\pi a) = N a_{\rm c}/a$ [see Eq.~(\ref{eq:dimcoef})]. This equilibrium stability criterion is not obviously applicable in the non-equilibrium situation after the first collapse~\cite{Cornish:2006}. Still, it provides a good estimate of both the point, $a_{\rm c}/|a| \approx - 0.6$, beyond which the single-collapse loss is insufficient to re-stabilise the system (see also Appendix C), and the long-time loss at large $a/a_{\rm c}$.

\begin{figure} [tbp]
\centering
\includegraphics[width=1\columnwidth]{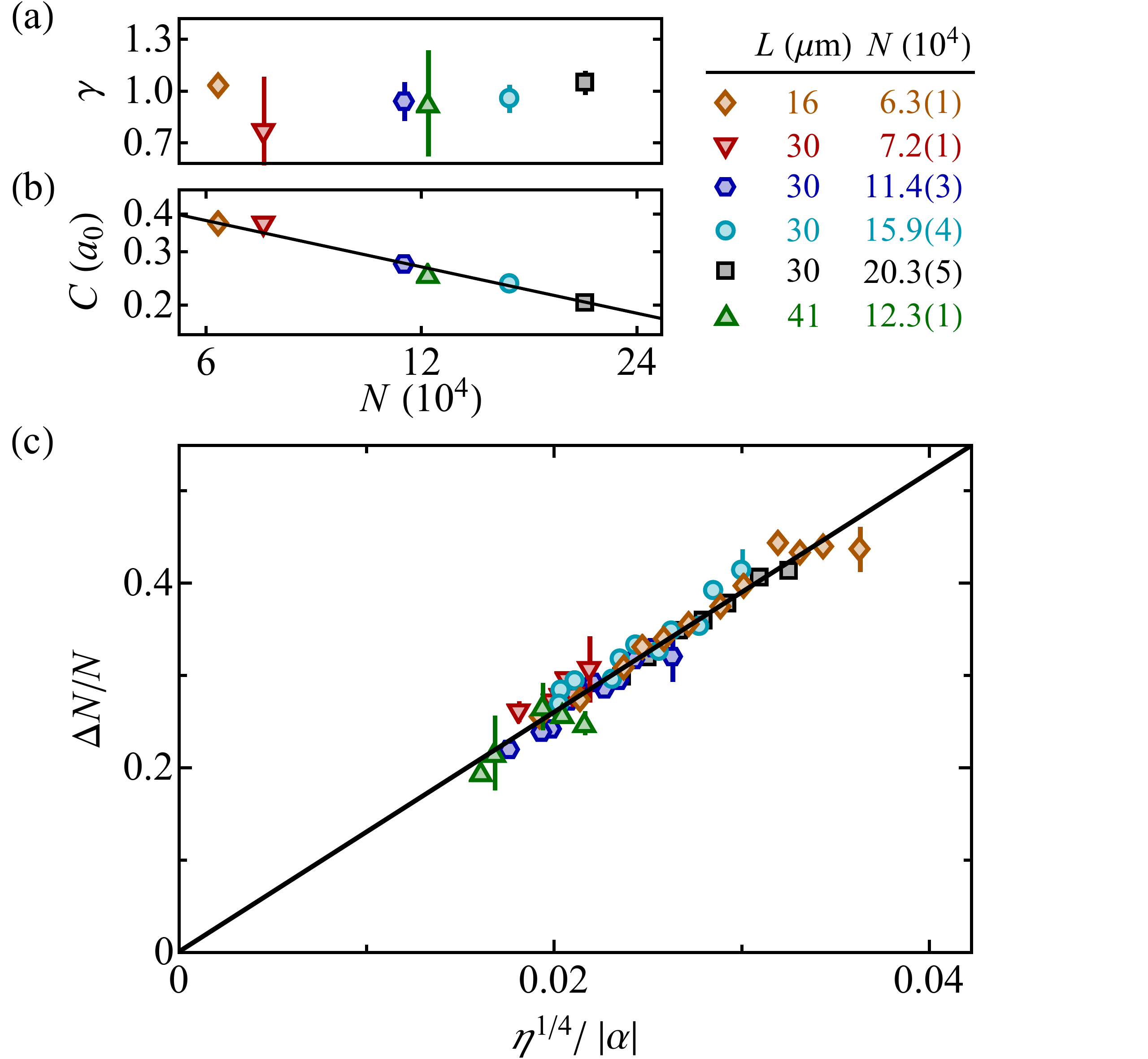}
\caption{Weak-collapse scaling laws.
(a) Writing $ \Delta N/N \propto |a|^{-\gamma}$ for fixed $L$ and $N$, for our six data sets we get an average $\bar{\gamma}=1.02(2)$.
(b) Assuming $\Delta N/N = C/|a|$, we find  $C \propto N^{-0.51(2)}$ (solid black line), with no dependence on $L$. (c) Universal behaviour of the weak-collapse atom loss. We plot all the single-collapse data for different $a$, $N$ and $L$ versus $\eta^{1/4}/|\alpha|$, which vanishes in the limit of infinitely strong attractive interactions, $|\alpha| \rightarrow \infty$. The linear fit gives $\Delta N/N  \approx 13\,   \eta^{1/4}/|\alpha|$. }
\label{fig:weakcollapse}
\end{figure}

We now extend the study of the weak-collapse atom loss to other $L$ and $N$ values (see Fig.~\ref{fig:weakcollapse}). In this analysis we include all $a$ values for which only single collapse occurs, and also those where clearly resolved single and double collapses occur (see Appendix C).

Writing $\Delta N/N \propto |a|^{-\gamma}$ for each data set with fixed $L$ and $N$, as in Fig.~\ref{fig:Nlost}(b),
we always get $\gamma$ consistent with unity [see Fig.~\ref{fig:weakcollapse}(a)]; averaging over all data sets gives $\bar{\gamma}=1.02(2)$.
We then assume the form $ \Delta N/N = C/|a|$ and study the dependence of $C$ on $L$ and $N$. As shown in Fig.~\ref{fig:weakcollapse}(b), on a log-log plot, we find  $C \propto N^{-0.51(2)}$, with no clear dependence on $L$; the two points taken with $L=16\,\mu$m and $41\,\mu$m fall onto the same line as the four points taken with $L=30\,\mu$m.

We thus experimentally find that weak-collapse atom loss is described very well by $\Delta N / N \propto 1/(\sqrt{N}|a|)$. From Eq.~(\ref{eq:dimcoef}), this corresponds to $\Delta N/N \propto \eta^{1/4}/|\alpha|$, which is indeed independent of $L$, and vanishes in the limit of infinitely strong attraction, $|\alpha| \rightarrow \infty$.
We note that while the weak collapse atom loss does not depend on $L$ (the overall size of the box) it may depend on the box shape; this is an interesting question for future research.

In Fig.~\ref{fig:weakcollapse}(c) we plot all our single-collapse data versus $\eta^{1/4}/|\alpha|$ and confirm that it falls onto a single universal curve~\cite{CollapseFootnote2}.
These experimentally obtained scaling laws should provide useful input for further theoretical work.

\section{Conclusions and outlook}
\label{sec:conclusion}

In conclusion,  we have performed a comprehensive study of the collapse of an attractive BEC confined in the homogeneous potential of a 3D box trap.
We have fully characterised the critical point for collapse, and the collapse dynamics of an interaction-quenched BEC, finding universal behaviour in agreement with the theoretical expectations.
Most importantly, we have provided conclusive experimental evidence for the counterintuitive weak collapse, and have experimentally determined weak-collapse scaling laws that should provide a useful reference point for the general theories of nonlinear wave phenomena.

Our work also points to many avenues for further research.  It would be very interesting to explore quenches from a large positive $a$, where the BEC is initially deep in the Thomas-Fermi regime, and in the case of a box potential the density is uniform. In this case it is not obvious how the condensate would spontaneously `choose' the position at which to collapse, or whether many local collapses would occur instead of a global one.
Additionally, since the fractional atom loss cannot exceed 100\%, the linear trend seen in Fig.~\ref{fig:weakcollapse}(c) cannot extend to the regime of strong dissipation (large $\eta$). It would be interesting to explore that regime using a different geometry, a different Feshbach resonance, or a different atomic species.
Finally, a major extension would be to perform similar experiments with 2D gases, for which a strong collapse and hence fundamentally different behaviour is expected.

\begin{acknowledgments}

We thank Sarah Thomas, Yago del Valle-Incl{\'a}n Redondo and Cornelius Roemer for experimental assistance, and Richard Fletcher, Raphael Lopes and Andreas Nunnenkamp for a critical reading of the manuscript. The GeForce GTX TITAN X used for the numerical simulations was donated by the NVIDIA Corporation.
This work was supported by the Royal Society, EPSRC [Grant No. EP/N011759/1], ERC (QBox), AFOSR and ARO. A.L.G. and N.N. acknowledge support from Trinity College, Cambridge.

\end{acknowledgments}

\section*{Appendix A - Experimental Setup }

Our setup is the first 3D BEC box experiment with tuneable interactions.
The setup for producing harmonically trapped $^{39}$K condensates is similar to our previous apparatus~\cite{Campbell:2010}. The main difference is that here we  employ  the gray molasses technique~\cite{Boiron:1996,Salomon:2013,Nath:2013} and directly cool $^{39}$K without the need for sympathetic cooling with rubidium atoms (see also~\cite{Landini:2012, Salomon:2014}). We load the laser-cooled atoms directly into a crossed optical dipole trap (using a $20$-W $1070$-nm laser) and achieve efficient evaporative cooling using the Feshbach resonance in the $|F,m_F\rangle = |1,1\rangle$ state at 402.70(3)~G~\cite{Fletcher:2016}. This results in a quasi-pure BEC of $\approx 2\times 10^5$ atoms.
We then load the atoms into a cylindrical optical box formed by blue-detuned ($532$~nm) laser light, and cancel out gravity with a magnetic field gradient, as in~\cite{Gaunt:2013}. The loading procedure is essentially 100\% efficient and results in a quasi-pure box-trapped BEC of  $\approx 2\times 10^5$ atoms.

The Feshbach resonance in the $|1,1\rangle$ state has a width of $\Delta B =52$~G and the background scattering length is $a_{\rm bg}=-29~a_0$ \cite{DErrico:2007}. Hence, near the zero-crossing of $a$, at $\approx 350$~G, the variation of the scattering length with the magnetic field is $-a_{\rm bg}/\Delta B \approx 0.6~a_0/$G. We tune $B$ in steps of 50~mG, corresponding to a scattering length resolution of 0.03~$a_0$. 

\section*{Appendix B - Scattering Length Calibration}

The exact magnetic field at which the scattering length in the $|1,1\rangle$ state vanishes was independently measured in Ref.~\cite{Fattori:2008b} to be \smash{$B_{a=0} = 350.4(1)$~G}.
For Fig.~\ref{fig:ac}(a-c) we calculate our $a$ values assuming \smash{$B_{a=0} = 350.4$~G}. Fitting the data in Fig.~\ref{fig:ac}(c) with a free intercept gives an intercept \smash{$a_{\rm c} (1/N=0) = 0.03(1)\,a_0$},  which is consistent with zero within the systematic $0.06\,a_0$ error due to the uncertainty in $B_{a=0}$.
We take this to be an unbiased confirmation of the zero intercept and the expected scaling $a_{\rm c} \propto 1/N$, and use this scaling to slightly refine the value of the zero-crossing field, to $B_{a=0} = 350.45(3)$~G.
The remaining 30~mG uncertainty in $B_{a=0}$ corresponds to a systematic uncertainty in our $a$ values of $\approx 0.02\,a_0$.

For our interaction quenches we have determined, using radio-frequency spectroscopy,  that the magnetic field takes 4~ms to change (from 20 to 80 \% of the jump). We account for this delay in our determination of the collapse time, and also include an additional 2~ms uncertainty in all the reported $t_{\rm c}$ values.

\section*{Appendix C - From Single to Double collapse}

\begin{figure} [b!]
\centering
\includegraphics[width=1\columnwidth]{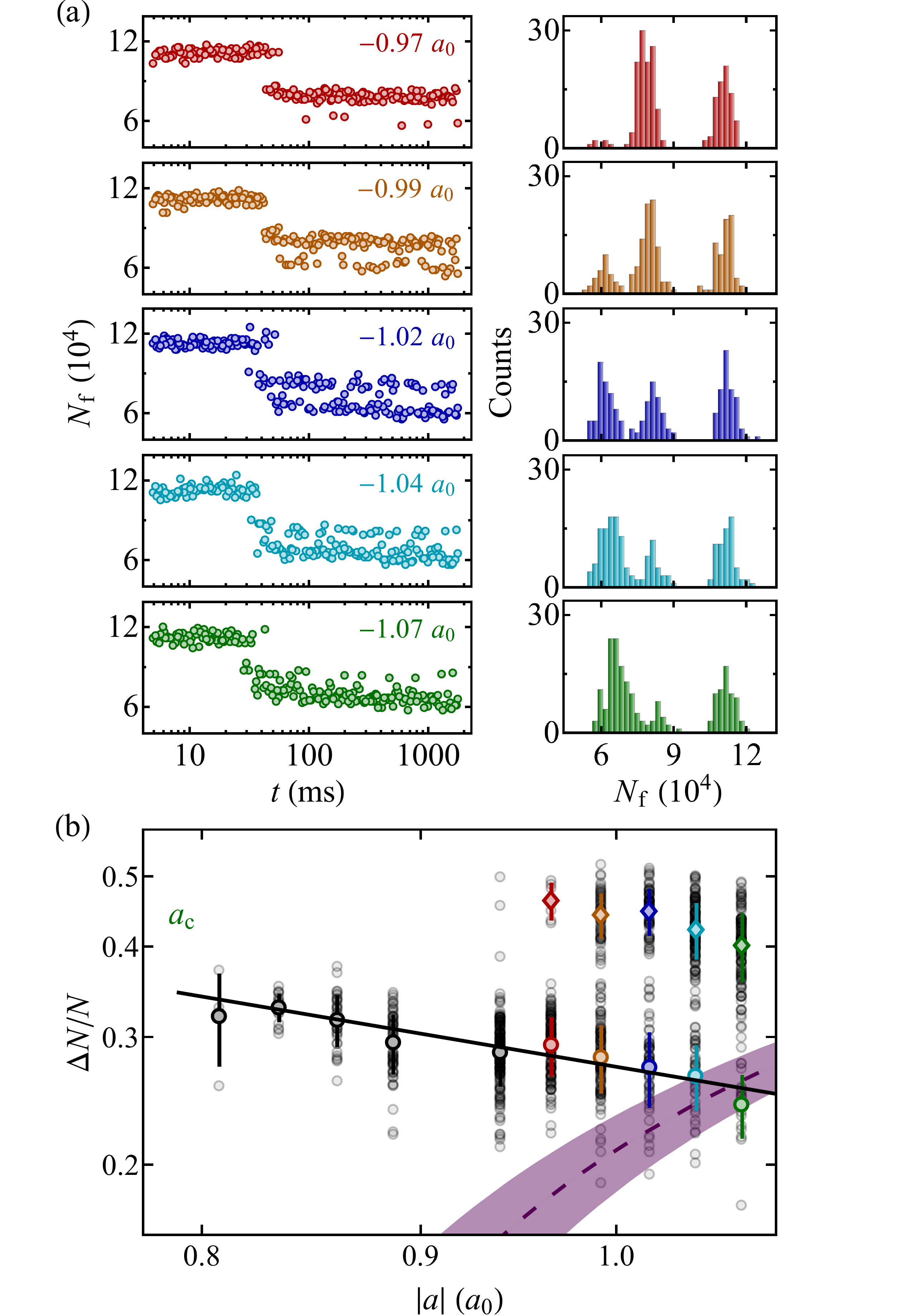}
\caption{Transition from one to two collapse events. (a) Atom number versus $t$ for $L=30~\mu$m,  initial $N=11.4 \times 10^4$,  and various closely spaced $a$ values; here $a_{\rm c} = -0.79\,a_0$. On the right we show histograms of $N_\text{f}$ values.  For $t>t_{\rm c}$ we see two clearly resolved $N_\text{f}$ branches, corresponding to one (upper branch) and two (lower branch) collapse events.
The probability of a double collapse gradually increases with $|a|$. (b) Fractional atom loss versus $|a|$ on a log-log plot. The transparent black circles show the raw data. The coloured circles and diamonds show, respectively, the average values for the single- and double-collapse events. The colour code is the same as in (a). The error bars show the standard deviations. The purple dashed line shows the BEC stability criterion as in Fig.~\ref{fig:Nlost}(b) and the shading shows its uncertainty.  The solid black line shows the (single-event) weak-collapse scaling, $\Delta N / N \propto 1/|a|$.
}
\label{fig:Bif}
\end{figure}

In Fig.~\ref{fig:Bif} we present  evidence for a gradual transition between single- and double-collapse events, which strongly supports the interpretation that an increasing number of discrete collapse events occur as $|a|$ is increased. This data was taken with $L=30~\mu$m and  initial $N=11.4 \times 10^4$.

In Fig.~\ref{fig:Bif}(a) we show the evolution of $N_\text{f}$ after a quench to various $a<a_\text{c}$. A fine scan of $a$ resolves a striking bifurcation of the collapse outcome. We interpret the upper and lower branch as the result of, respectively, one and two collapse events. As $|a|$ is increased, the probability of a double collapse gradually increases.
This crossover is highlighted in the histograms shown on the right.

In Fig.~\ref{fig:Bif}(b) we show the fractional atom loss versus $|a|$ on a log-log plot.
In the regime where a double collapse occurs, the single-collapse branch still clearly follows the weak-collapse scaling $\Delta N / N \propto 1/|a|$.
Note that in this data set the double collapse occurs slightly closer to $a_\text{c}$ than expected from the simple equilibrium stability criterion (purple band).

\newpage


\begin{thebibliography}{61}
\bibitem [{\citenamefont {{}}()}]{Definetc}%
  \BibitemOpen
  \href@noop {}
  {}\bibinfo {note} {{N}ote that in general $t_{\rm c}$ is best defined as the time when the maximum density is reached and the atom loss rate is the highest. In all our data sets we observe the highest loss rate at the onset of atom loss. }\BibitemShut {Stop}%
\bibitem [{\citenamefont {{}}()}]{CollapseFootnote2}%
  \BibitemOpen
  \href@noop {}
  {}\bibinfo {note} {{H}ere each point and its error bar show the average and
  standard error for a cluster of single-collapse points taken for same
  $\{a,N,L\}$ and just different $t>t_{\rm c}$, such as seen in
  Fig.~\ref{fig:Nlost}(b) for $a_{\rm c}/|a| < - 0.6$.}\BibitemShut {Stop}%
\end{thebibliography}


%


\end{document}